\documentclass[twocolumn,showpacs,preprintnumbers,amsmath,amssymb]{revtex4}

\usepackage{graphicx}
\usepackage{dcolumn}
\usepackage{bm}

\newcommand{\kagome}{ {Kagom\'e} }

\begin{document}

\title{Hidden long range order in Heisenberg \kagome antiferromagnets}

\author{A.V. Syromyatnikov}
 \email{syromyat@thd.pnpi.spb.ru}
\author{S.V. Maleyev}
\affiliation{Petersburg Nuclear Physics Institute, Gatchina, St.\ Petersburg 188300, Russia}

\date{\today}

\begin{abstract}

We give a physical picture of the low-energy sector of the spin 1/2 Heisenberg \kagome antiferromagnet (KAF). It
is shown that Kagom\'e lattice can be presented as a set of stars which are arranged in a triangular lattice and
contain 12 spins. Each of these stars has two degenerate singlet ground states which can be considered in terms
of pseudospin. As a result of interaction between stars we get Hamiltonian of the Ising ferromagnet in magnetic
field. So in contrast to the common view there is a long range order in KAF consisting of definite singlet
states of the stars.

\end{abstract}


\maketitle

In spite of numerous theoretical and experimental studies in the last decade, some magnetic properties of
\kagome antiferromagnets (KAFs) remain open problems. Experiments revealed unusual low-temperature behavior of
the specific heat and magnetic susceptibility in \kagome-like compounds. For example specific heat $C$
measurements in $SrCr_{9p}Ga_{12-9p}O_{19}$ ($S=3/2$ \kagome material) have shown that there is a peak at
$T\approx 5$ K, $C\propto T^2$ at $T\alt 5$ K and it appears to be practically independent of magnetic field up
to 12 T \cite{ramirez}.

Some of the experimental findings are in agreement with the results of numerical finite cluster investigations
\cite{zeng,leung,wald,zeng2}. They reveal a gap separating the ground state from the upper triplet levels and a
band of nonmagnetic singlet excitations with a very small or zero gap inside the triplet gap. The number of
states in the band increases with the number of sites $N$ as $\alpha^N$. For samples with up to 36 sites
$\alpha=1.15$ and 1.18 for the even and odd $N$, respectively \cite{wald}. This wealth of low-lying singlet
excitations explains the peak of specific heat at low temperature and its independence of the magnetic field.

However the origin of this band as well as the nature of the ground state were unclear until now. Previous exact
diagonalization studies \cite{chalker,leung} reveal exponential decay of the spin-spin and dimer-dimer
correlation functions. So the point of view that KAF is a spin liquid is widely accepted now
\cite{sachdev,zeng2,yang,kalmeyer,marston,wald,leung,chalker}. It seems the best candidate for description of
KAF low-energy properties is a quantum dimer model \cite{sachdev,zeng2,mila1}. It should be mentioned a certain
recent success in this field. In the paper \cite{mila2} a spin 1/2 \kagome lattice is considered as a set of
interactive triangles with a spin in each apex. It was suggested there to work in the subspace where the total
spin of each triangle is 1/2 (short range RVB states (SRRVB)) investigating low-lying excitations. It was shown
that low-energy spectrum obtained with SRRVB on the samples with up to 36 cites and the number of singlet
excitations in the band coincide with the results of exact diagonalization. Meanwhile a further development of
this approach is required to get a full physical description of KAF.

Another types of frustrated magnets which possess a similar behavior as KAF and have many singlet states inside
the triplet gap are pyrochlore \cite{canals} and $CaV_4O_9$ \cite{alb}. Recently it was suggested a model of
frustrated antiferromagnet which low-energy properties can be generic for these compounds as well as for KAF
\cite{zhit}. Weakly interactive plaquettes in the square lattice were considered there. Each plaquette has two
almost degenerate singlet ground states, so a band of singlet excitations arises if the inter-plaquette
interaction is taken into account. It is shown that there is a quantum phase transition in the model at a
critical value of frustration separating a disorder plaquette phase and a columnar dimer one. In the proximity
of this transition the specific heat has a low-temperature peak below which it possesses a power low temperature
dependence.

\begin{figure}
  \centering
  \includegraphics{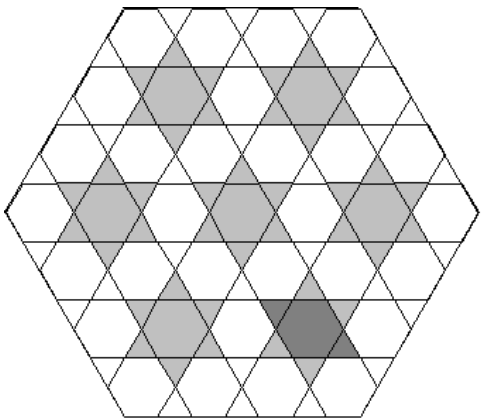}
  \caption{
\kagome lattice (KL). There is a spin in each lattice site. KL can be considered as a set of stars arranged in a
triangular lattice. Each star contains 12 spins. An unit cell is also presented (dark region). There are four
unit cells per a star.
  \label{lattice}}
\end{figure}

In this paper we show that such a behavior is do relevant for spin 1/2 KAF. It is proposed to consider a \kagome
lattice as a set of stars with 12 spins arranged in a triangular lattice (see Fig.~\ref{lattice}). A star has
two degenerate singlet ground states. Interaction between stars leads to the band of low-lying excitations which
number increases as $2^{N/12}\approx1.06^N$. It is demonstrated that this interaction can be considered as a
perturbation in the low-energy sector. As a result we get a model of the Ising ferromagnet in effective magnetic
field where these degenerate states are described in terms of two projections of pseudospins 1/2. So it is shown
that in contrast to the common view there is a hidden long range order in KAF which consists of definite singlet
states of the stars. This picture should be relevant also for KAFs with larger values of spin.

We start with the Hamiltonian of the spin 1/2 \kagome Heisenberg antiferromagnet:
\begin{equation}
{\cal H}_0 = J_1\sum_{\langle i,j \rangle}{\bf S}_i{\bf S}_j + J_2\sum_{(i,j)}{\bf S}_i{\bf S}_j, \label{h}
\end{equation}
where $\langle i,j \rangle$ and $(i,j)$ denote nearest and next-nearest neighbors on the \kagome lattice shown
in Fig.~\ref{lattice}, respectively. The case of $|J_2|\ll J_1$ is considered in this paper. We discuss a
possibility of both signs of next-nearest-neighbor interaction --- ferromagnet and antiferromagnet one. As is
shown below, in spite of the smallness the second term in Eq.~(\ref{h}) can be of importance for the low-energy
properties.

\begin{figure}
  \centering
  \includegraphics{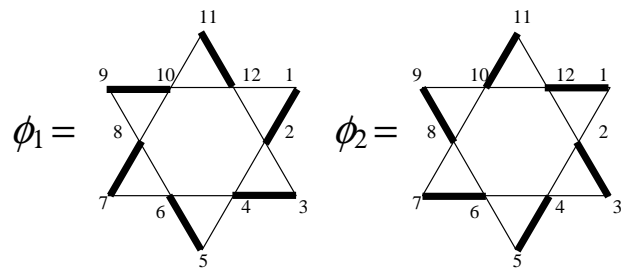}
  \caption{
Schematic representation of a star's two singlet ground states wave functions $\phi_1$ and $\phi_2$.  A bold
line denotes the singlet state of two neighboring spins, i.e. $({\mid\uparrow\rangle}_i
{\mid\downarrow\rangle}_j - {\mid\downarrow\rangle}_i{\mid\uparrow\rangle}_j)/\sqrt{2}$.
  \label{grounds}}
\end{figure}

\kagome lattice can be presented as a set of stars arranged in a triangular lattice (see Fig.~\ref{lattice}). To
begin with we neglect interaction between stars and put $J_2=0$ in Eq.~(\ref{h}). A star is a system of 12
spins. Let us consider its properties in detail. The symmetry group contains 6 rotations and reflections with
respect to 6 lines passing through the center. There are two degenerate singlet ground states $\phi_1$ and
$\phi_2$ which differ each other by symmetry. Their wave functions are shown schematically in Fig.~\ref{grounds}
where a bold line represents the singlet state of the corresponding two spins, i.e. $({\mid\uparrow\rangle}_i
{\mid\downarrow\rangle}_j - {\mid\downarrow\rangle}_i{\mid\uparrow\rangle}_j)/\sqrt{2}$. Evidently $\phi_1$ and
$\phi_2$ are invariant with respect to rotations of the star and they transform to each other under reflections.
They contain six singlets each having the energy $-3/4J_1$. One can show that interaction between singlets makes
no contribution to the energy of the ground states which is consequently equal to $-4.5J_1$. We have obtained
numerically that there is a gap of the value approximately $0.26J_1$ which separates the ground states and the
lower triplet levels in the star.

It should be pointed out that $\phi_1$ and $\phi_2$ are not orthogonal: the scalar product of these two
functions is $(\phi_1\phi_2)=1/32$. So in the following we will use two orthonormalized combinations:
\begin{eqnarray}
\Psi_1&=&\frac{1}{\sqrt{2+1/16}}(\phi_1+\phi_2), \label{f1}\\
\Psi_2&=&\frac{1}{\sqrt{2-1/16}}(\phi_1-\phi_2). \label{f2}
\end{eqnarray}
Because $\phi_1$ and $\phi_2$ are invariant under rotations and transform to each other with reflections,
$\Psi_1$ is invariant under all the symmetry group transformations and $\Psi_2$ is invariant under rotations and
it changes the sign with reflections.

\begin{figure}
  \centering
  \includegraphics{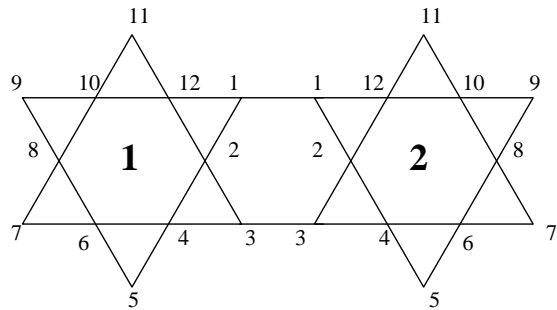}
  \caption{
There are the following interactions between each two stars: $V=J_1({\bf S}^{(1)}_1{\bf S}^{(2)}_1+{\bf
S}^{(1)}_3{\bf S}^{(2)}_3)$ and $\tilde V_1=J_2({\bf S}^{(1)}_1{\bf S}^{(2)}_2+{\bf S}^{(1)}_2{\bf
S}^{(2)}_1+{\bf S}^{(1)}_2{\bf S}^{(2)}_3+{\bf S}^{(1)}_3{\bf S}^{(2)}_2)$, where upper indexes label the stars.
  \label{twostars}}
\end{figure}

We turn now to consideration of the interaction between two nearest stars still neglecting the second term in
Eq.~(\ref{h}). Initially there are four degenerate ground states with energy $E_0=-9J_1$ and wave functions $\{
\Psi^{(1)}_{n_1}\Psi^{(2)}_{n_2} \}$ ($n_i=1,2$), where upper index labels the stars. As it is seen from
Fig.~\ref{twostars}, the interaction energy has the form:
\begin{equation}\label{v}
V=J_1({\bf S}^{(1)}_1{\bf S}^{(2)}_1+{\bf S}^{(1)}_3{\bf S}^{(2)}_3).
\end{equation}
Let us consider $V$ as a perturbation. According to the standard theory
\cite{landau} one have to solve a secular equation to find the first correction to the energy. In our case there
are four equations and the corresponding matrix elements are given by
\begin{equation}\label{siecle}
    H_{n_1n_2;k_1k_2} = V_{n_1n_2;k_1k_2}+\sum_{m_1,m_2}\frac{V_{n_1n_2;m_1m_2}V_{m_1m_2;k_1k_2}}
    {E_0-E_{m_1}-E_{m_2}},
\end{equation}
where $V_{n_1n_2;k_1k_2}=\langle \Psi^{(1)}_{n_1}\Psi^{(2)}_{n_2}\mid V
\mid\Psi^{(1)}_{k_1}\Psi^{(2)}_{k_2}\rangle$, $n_i,k_i=1,2$ and indexes $m_1$ and $m_2$ denote excited levels of
the first and the second star, respectively. Obviously the first term in Eq.~(\ref{siecle}) is zero and the
second one can be represented as follows:
\begin{eqnarray}\label{s2}
    H_{n_1n_2;k_1k_2}&=&-i\int_0^\infty dt e^{-\delta t+iE_0t}\nonumber\\
    &&\langle \Psi^{(1)}_{n_1}\Psi^{(2)}_{n_2}\mid
    Ve^{-it(H_{01}+H_{02})}V\mid\Psi^{(1)}_{k_1}\Psi^{(2)}_{k_2}\rangle,\nonumber\\
\end{eqnarray}
where $H_{0i}$ are Hamiltonians of the corresponding stars. Using the symmetry of the functions $\phi_1$,
$\phi_2$, $\Psi_1$ and $\Psi_2$ discussed above one can show that only diagonal elements (i.e.\ $n_1=k_1$,
$n_2=k_2$) in Eq.~(\ref{s2}) are nonzero. We have calculated them with a very high precision by expansion of the
operator exponent up to the power 130. The results can be represented in the following form:
\begin{subequations}\label{mel}
\begin{eqnarray}
H_{11;11}&=&-a_1+a_2-a_3,\\
H_{12;12}&=&-a_1+a_3,\\
H_{21;21}&=&-a_1+a_3,\\
H_{22;22}&=&-a_1-a_2-a_3,
\end{eqnarray}
\end{subequations}
where $a_1=0.256J_1$, $a_2=0.015J_1$ and $a_3=0.0027J_1$. So the interaction shifts all the levels on the value
$-a_1$ and lifts their degeneracy. Constants $a_2$ and $a_3$ in Eqs.~(\ref{mel}) determine the levels splitting.
All corrections are small enough and one can consider interaction Eq.~(\ref{v}) between stars as a perturbation
at low energies. We restrict ourself with this precision here and don't consider triplet states.

So KAF appears to be a set of two-levels interacting systems and one can naturally represent the low-energy
singlet sector of Hilbert space in terms of pseudospins: ${\mid\uparrow\rangle}=\Psi_2$ and
${\mid\downarrow\rangle}=\Psi_1$. It is seen from Eqs.~(\ref{mel}) that in these terms the interaction between
stars is described by the Hamiltonian of Ising ferromagnet in the external magnetic field:
\begin{equation}\label{ham}
    {\cal H}=-{\cal J}\sum_{\langle i,j\rangle}s_i^zs_j^z-h\sum_is_i^z ,
\end{equation}
where $\langle i,j\rangle$ labes now nearest-neighbor pseudospins, arranged in a triangular lattice formed by
the stars, ${\cal J}=4a_3=0.011J_1$ and $h=6a_2=0.092J_1$. We also omit a constant in Eq.~(\ref{ham}) which is
equal to $-0.439J_1N$. It should be stressed that within our precision Hamiltonian Eq.~(\ref{ham}) is an exact
mapping of the original Heisenberg model in the low-energy sector (excitation energy $\omega\sim{\cal J}$). So
one can see from Eq.~(\ref{ham}) that the ground state of KAF is that with all the stars in $\Psi_2$ states.

In fact we show existence of a long range order in KAF generated by singlets. This hidden order settles on the
triangular star lattice and can be checked by inelastic neutron scattering: corresponding intensity for the
singlet-triplet transitions should have periodicity in the reciprocal space corresponding to the original star
lattice. This picture is similar to observed one in the case of the dimerised spin-Pairls compound $CuGeO_3$
\cite{germ}. Because low-energy physics in KAF is determined by singlets, our consideration should be relevant
qualitatively also for KAFs with the larger values of spin.

We proceed with the discussion of the number of low-energy states in KAF. As each star has two singlet ground
states and contains 12 spins the number of singlet excitations in the band is given by $2^{N/12}\approx 1.06^N$.
Unfortunately there is no point to compare this result with that of the previous numerical work \cite{wald}
discussed above because a very small samples ($N\le36$) were considered there.

Let us take into account the next-nearest-neighbor interactions. As is seen from Fig.~\ref{twostars} they can be
divided on three parts: $\tilde V_1$, $\tilde V^{(1)}_2$ and $\tilde V^{(2)}_2$, where $\tilde V_1=J_2({\bf
S}^{(1)}_1{\bf S}^{(2)}_2+{\bf S}^{(1)}_2{\bf S}^{(2)}_1+{\bf S}^{(1)}_2{\bf S}^{(2)}_3+{\bf S}^{(1)}_3{\bf
S}^{(2)}_2)$ contributes to the inter-stars interaction and $\tilde V^{(1)}_2$ and $\tilde V^{(2)}_2$ contain 12
intrinsic next-nearest-neighbor interactions of the first and the second star, respectively. Considering now
perturbation theory according to a sum of these three operators and that given by Eq.~(\ref{v}) we find that in
addition to corrections presented in Eqs.~(\ref{mel}) there are new ones proportional to $J_2$ from the first
and the second terms in Eq.~(\ref{siecle}) given by $\tilde V^{(1)}_2$, $\tilde V^{(2)}_2$ and $\tilde V_1$,
respectively. Using symmetry of functions $\phi_1$ and $\phi_2$ it can be shown that secular matrix
Eq.~(\ref{s2}) remains diagonal in this case. As a result calculations give for the values of "exchange" and
"magnetic field" in the effective Hamiltonian Eq.~(\ref{ham}):
\begin{eqnarray}
{\cal J}&=&0.011J_1-0.005J_2,\label{j}\\
h&=&0.092J_1+3.635J_2.\label{magf}
\end{eqnarray}
One can see from Eqs.~(\ref{j}) and (\ref{magf}) that the next-nearest interactions give a correction of the
order of $|J_2|/J_1\ll1$ to the value $\cal J$ and their contribution to the "magnetic field" is considerable if
$|J_2|\agt 0.01J_1$. If $J_2<0$ (ferromagnet interaction) they could even change the sign of $h$. In the case of
$h>0$ the ground state of the \kagome lattice is that with all stars in $\Psi_2$ states and if $h<0$ all stars
are in $\Psi_1$ states.

We could expect a logarithmic singularity of the specific heat $C$ in the point $h=0$ at the critical
temperature $T_c$ which is of the order of $\cal J$ and there should be a peak at $T\sim \cal J$ if $h\ne 0$.
Specific heat decreases at $T\to0$ as $e^{-(3{\cal J}+|h|)/T}$. So we don't get low-temperature behavior
$C\propto T^2$ obtained in experiments \cite{ramirez}. It should be noted that such a law should exist if the
energy of low-lying excitations $\epsilon_{\bf q}$ with wave vector $\bf q$ at $q\ll1$ has the form
$\epsilon_{\bf q}=cq^2+\Delta$. Within the first order of perturbation theory considered in this paper
interaction between stars is described by the Hamiltonian Eq.~(\ref{ham}) of Ising ferromagnet in magnetic field
which doesn't imply such a behavior of the low-energy spectrum. But the further orders could give some kind of
anisotropy in Eq.~(\ref{ham}) which leads to the necessary picture. This point will be considered in detail
elsewhere.

It is appropriate to mention here a recent experiment on $Cu_3V_2O_7(OH)_2\cdot 2H_2O$ \cite{hiroi} which is the
only candidate for spin-1/2 \kagome material by now. In spite of strong exchange $J_1\sim 100$ K in this
compound, there is no regime obtained for KAF with the larger values of spins has been achieved up to the
temperature $1.8$ K. In this respect we point here on a small scale of the dynamics in spin-1/2 KAF. According
to our results the representative temperature is of the order of $0.01J_1$, so the region $T\alt 1$ K should be
attained for this material.

In conclusion, we present a new insight of low-energy physics of spin 1/2 \kagome antiferromagnet (KAF). The
lattice can be presented as a set of stars which are arranged in a triangular lattice and contain 12 spins (see
Fig.~\ref{lattice}). Each star has two degenerate singlet ground states with different symmetry. It is shown
that interaction between the stars leads to the band of singlet excitations and can be considered as a
perturbation in the low-energy sector. We demonstrate the existence of a long range order in KAF on the
triangular star lattice which is generated by singlets and can be detected in particular in experiments on
inelastic neutron scattering. This physical picture should be relevant also for KAFs with larger spin values.

This work was supported by Russian State Program "Collective and Quantum Effects in Condensed Matter",
the Russian Foundation for Basic Research (Grants No.\ 00-02-16873, 00-15-96814) and Russian State
Program "Quantum Macrophysics".

\bibliography{kagome}

\begin{thebibliography}{18}
\expandafter\ifx\csname natexlab\endcsname\relax\def\natexlab#1{#1}\fi
\expandafter\ifx\csname bibnamefont\endcsname\relax
  \def\bibnamefont#1{#1}\fi
\expandafter\ifx\csname bibfnamefont\endcsname\relax
  \def\bibfnamefont#1{#1}\fi
\expandafter\ifx\csname citenamefont\endcsname\relax
  \def\citenamefont#1{#1}\fi
\expandafter\ifx\csname url\endcsname\relax
  \def\url#1{\texttt{#1}}\fi
\expandafter\ifx\csname urlprefix\endcsname\relax\def\urlprefix{URL }\fi
\providecommand{\bibinfo}[2]{#2}
\providecommand{\eprint}[2][]{\url{#2}}

\bibitem[{\citenamefont{Ramirez et~al.}(2000)\citenamefont{Ramirez, Hessen, and
  Winclemann}}]{ramirez}
\bibinfo{author}{\bibfnamefont{A.~P.} \bibnamefont{Ramirez}},
  \bibinfo{author}{\bibfnamefont{B.}~\bibnamefont{Hessen}}, \bibnamefont{and}
  \bibinfo{author}{\bibfnamefont{M.}~\bibnamefont{Winclemann}},
  \bibinfo{journal}{Phys.\ Rev.\ Lett.} \textbf{\bibinfo{volume}{84}},
  \bibinfo{pages}{2957} (\bibinfo{year}{2000}).

\bibitem[{\citenamefont{Zeng and Elser}(1990)}]{zeng}
\bibinfo{author}{\bibfnamefont{C.}~\bibnamefont{Zeng}} \bibnamefont{and}
  \bibinfo{author}{\bibfnamefont{V.}~\bibnamefont{Elser}},
  \bibinfo{journal}{Phys.\ Rev.\ B} \textbf{\bibinfo{volume}{42}},
  \bibinfo{pages}{8436} (\bibinfo{year}{1990}).

\bibitem[{\citenamefont{Waldtmann et~al.}(1998)\citenamefont{Waldtmann, Everts,
  Bernu, Sindzingre, Lhuillier, Lecheminant, and Pierre}}]{wald}
\bibinfo{author}{\bibfnamefont{C.}~\bibnamefont{Waldtmann}},
  \bibinfo{author}{\bibfnamefont{H.-U.} \bibnamefont{Everts}},
  \bibinfo{author}{\bibfnamefont{B.}~\bibnamefont{Bernu}},
  \bibinfo{author}{\bibfnamefont{P.}~\bibnamefont{Sindzingre}},
  \bibinfo{author}{\bibfnamefont{C.}~\bibnamefont{Lhuillier}},
  \bibinfo{author}{\bibfnamefont{P.}~\bibnamefont{Lecheminant}},
  \bibnamefont{and} \bibinfo{author}{\bibfnamefont{L.}~\bibnamefont{Pierre}},
  \bibinfo{journal}{Eur.\ Phys.\ J.\ B} \textbf{\bibinfo{volume}{2}},
  \bibinfo{pages}{501} (\bibinfo{year}{1998}).

\bibitem[{\citenamefont{Leung and Elser}(1993)}]{leung}
\bibinfo{author}{\bibfnamefont{P.~W.} \bibnamefont{Leung}} \bibnamefont{and}
  \bibinfo{author}{\bibfnamefont{V.}~\bibnamefont{Elser}},
  \bibinfo{journal}{Phys.\ Rev.\ B} \textbf{\bibinfo{volume}{47}},
  \bibinfo{pages}{5459} (\bibinfo{year}{1993}).

\bibitem[{\citenamefont{Zeng and Elser}(1995)}]{zeng2}
\bibinfo{author}{\bibfnamefont{C.}~\bibnamefont{Zeng}} \bibnamefont{and}
  \bibinfo{author}{\bibfnamefont{V.}~\bibnamefont{Elser}},
  \bibinfo{journal}{Phys.\ Rev.\ B} \textbf{\bibinfo{volume}{51}},
  \bibinfo{pages}{8318} (\bibinfo{year}{1995}).

\bibitem[{\citenamefont{Chalker and Eastmond}(1992)}]{chalker}
\bibinfo{author}{\bibfnamefont{J.~T.} \bibnamefont{Chalker}} \bibnamefont{and}
  \bibinfo{author}{\bibfnamefont{J.~F.~G.} \bibnamefont{Eastmond}},
  \bibinfo{journal}{Phys.\ Rev.\ B} \textbf{\bibinfo{volume}{46}},
  \bibinfo{pages}{14201} (\bibinfo{year}{1992}).

\bibitem[{\citenamefont{Sachdev}(1992)}]{sachdev}
\bibinfo{author}{\bibfnamefont{S.}~\bibnamefont{Sachdev}},
  \bibinfo{journal}{Phys.\ Rev. B} \textbf{\bibinfo{volume}{45}},
  \bibinfo{pages}{12377} (\bibinfo{year}{1992}).

\bibitem[{\citenamefont{Yang et~al.}(1993)\citenamefont{Yang, Warman, and
  Girvin}}]{yang}
\bibinfo{author}{\bibfnamefont{K.}~\bibnamefont{Yang}},
  \bibinfo{author}{\bibfnamefont{L.~K.} \bibnamefont{Warman}},
  \bibnamefont{and} \bibinfo{author}{\bibfnamefont{S.~M.}
  \bibnamefont{Girvin}}, \bibinfo{journal}{Phys.\ Rev.\ Lett.}
  \textbf{\bibinfo{volume}{70}}, \bibinfo{pages}{2641} (\bibinfo{year}{1993}).

\bibitem[{\citenamefont{Kalmeyer and Laughlin}(1987)}]{kalmeyer}
\bibinfo{author}{\bibfnamefont{V.}~\bibnamefont{Kalmeyer}} \bibnamefont{and}
  \bibinfo{author}{\bibfnamefont{R.~B.} \bibnamefont{Laughlin}},
  \bibinfo{journal}{Phys.\ Rev.\ Lett.} \textbf{\bibinfo{volume}{59}},
  \bibinfo{pages}{2095} (\bibinfo{year}{1987}).

\bibitem[{\citenamefont{Marston and Zeng}(1991)}]{marston}
\bibinfo{author}{\bibfnamefont{J.~B.} \bibnamefont{Marston}} \bibnamefont{and}
  \bibinfo{author}{\bibfnamefont{C.}~\bibnamefont{Zeng}}, \bibinfo{journal}{J.\
  Appl.\ Phys.} \textbf{\bibinfo{volume}{69}}, \bibinfo{pages}{5962}
  (\bibinfo{year}{1991}).

\bibitem[{\citenamefont{Mila}(1998)}]{mila1}
\bibinfo{author}{\bibfnamefont{F.}~\bibnamefont{Mila}},
  \bibinfo{journal}{Phys.\ Rev.\ Lett.} \textbf{\bibinfo{volume}{81}},
  \bibinfo{pages}{2356} (\bibinfo{year}{1998}).

\bibitem[{\citenamefont{Mambrini and Mila}()}]{mila2}
\bibinfo{author}{\bibfnamefont{M.}~\bibnamefont{Mambrini}} \bibnamefont{and}
  \bibinfo{author}{\bibfnamefont{F.}~\bibnamefont{Mila}},
  \eprint{cond-mat/0003080}.

\bibitem[{\citenamefont{Canals and Lacroix}(1998)}]{canals}
\bibinfo{author}{\bibfnamefont{B.}~\bibnamefont{Canals}} \bibnamefont{and}
  \bibinfo{author}{\bibfnamefont{C.}~\bibnamefont{Lacroix}},
  \bibinfo{journal}{Phys.\ Rev.\ Lett.} \textbf{\bibinfo{volume}{80}},
  \bibinfo{pages}{2933} (\bibinfo{year}{1998}).

\bibitem[{\citenamefont{Albrecht et~al.}(1996)\citenamefont{Albrecht, Mila, and
  Poilblanc}}]{alb}
\bibinfo{author}{\bibfnamefont{M.}~\bibnamefont{Albrecht}},
  \bibinfo{author}{\bibfnamefont{F.}~\bibnamefont{Mila}}, \bibnamefont{and}
  \bibinfo{author}{\bibfnamefont{D.}~\bibnamefont{Poilblanc}},
  \bibinfo{journal}{Phys.\ Rev. B} \textbf{\bibinfo{volume}{54}},
  \bibinfo{pages}{15 856} (\bibinfo{year}{1996}).

\bibitem[{\citenamefont{Kotov et~al.}(2001)\citenamefont{Kotov, Zhitomirsky,
  and Sushkov}}]{zhit}
\bibinfo{author}{\bibfnamefont{V.~N.} \bibnamefont{Kotov}},
  \bibinfo{author}{\bibfnamefont{M.~E.} \bibnamefont{Zhitomirsky}},
  \bibnamefont{and} \bibinfo{author}{\bibfnamefont{O.~P.}
  \bibnamefont{Sushkov}}, \bibinfo{journal}{Phys.\ Rev.\ B}
  \textbf{\bibinfo{volume}{63}}, \bibinfo{pages}{064412}
  (\bibinfo{year}{2001}).

\bibitem[{\citenamefont{Landau and Livshitz}()}]{landau}
\bibinfo{author}{\bibfnamefont{L.~D.} \bibnamefont{Landau}} \bibnamefont{and}
  \bibinfo{author}{\bibfnamefont{E.~M.} \bibnamefont{Livshitz}},
  \eprint{Quantum mechaniks, Pergamon, Oxford, 1977}.

\bibitem[{\citenamefont{Regnault et~al.}(1996)\citenamefont{Regnault, Ain,
  Hennion, Dhalenne, and Revcolevschi}}]{germ}
\bibinfo{author}{\bibfnamefont{L.~P.} \bibnamefont{Regnault}},
  \bibinfo{author}{\bibfnamefont{M.}~\bibnamefont{Ain}},
  \bibinfo{author}{\bibfnamefont{B.}~\bibnamefont{Hennion}},
  \bibinfo{author}{\bibfnamefont{G.}~\bibnamefont{Dhalenne}}, \bibnamefont{and}
  \bibinfo{author}{\bibfnamefont{A.}~\bibnamefont{Revcolevschi}},
  \bibinfo{journal}{Phys.\ Rev.\ B} \textbf{\bibinfo{volume}{53}},
  \bibinfo{pages}{5579} (\bibinfo{year}{1996}).

\bibitem[{\citenamefont{Hiroi et~al.}()\citenamefont{Hiroi, Hanawa, Kobayashi,
  Nohara, Takagi, Kato, and Takigawa}}]{hiroi}
\bibinfo{author}{\bibfnamefont{Z.}~\bibnamefont{Hiroi}},
  \bibinfo{author}{\bibfnamefont{M.}~\bibnamefont{Hanawa}},
  \bibinfo{author}{\bibfnamefont{N.}~\bibnamefont{Kobayashi}},
  \bibinfo{author}{\bibfnamefont{M.}~\bibnamefont{Nohara}},
  \bibinfo{author}{\bibfnamefont{H.}~\bibnamefont{Takagi}},
  \bibinfo{author}{\bibfnamefont{Y.}~\bibnamefont{Kato}}, \bibnamefont{and}
  \bibinfo{author}{\bibfnamefont{M.}~\bibnamefont{Takigawa}},
  \eprint{cond-mat/0111127}.

\end{thebibliography}

\end{document}